\documentclass[12pt]{article}
\usepackage{amsfonts}
\usepackage{amsmath, amsthm}
\usepackage{epsfig}
\usepackage{algorithm}
\usepackage{algorithmic}
\usepackage{epic}
\usepackage[a4paper, bottom = 1.50cm, top = 2.0cm, left = 2.5cm
        ]{geometry}
\usepackage{enumitem}

\DeclareMathOperator*{\argmin}{arg\,min}

\usepackage{amsmath,verbatim,color,amssymb,epsfig}
\usepackage{bm}
\usepackage{ragged2e}
\usepackage{amsfonts}
\usepackage{epsfig}
\usepackage{changepage}
\usepackage{multirow}
\usepackage{graphicx}
\usepackage{array}
\usepackage[table]{xcolor}
\definecolor{Gray}{gray}{.80}
\usepackage{lscape}
\usepackage[round,semicolon,authoryear]{natbib}
\usepackage[normalem]{ulem}
\usepackage[colorlinks=true,urlcolor=blue,citecolor=purple,linkcolor=blue,bookmarks=true]{hyperref}
\usepackage{threeparttable}
\usepackage{caption}
\usepackage{pdfsync}
\usepackage{array}

\usepackage{booktabs}  
\usepackage{cleveref} 
\usepackage{caption}  
\usepackage{tabularx}   
\usepackage{adjustbox} 
\usepackage{longtable}  

\begin{document}
\def\eqx"#1"{{\label{#1}}}
\def\eqn"#1"{{\ref{#1}}}

\makeatletter 
\@addtoreset{equation}{section}
\makeatother  

\def\yuancomment#1{\vskip 2mm\boxit{\vskip 2mm{\color{red}\bf#1} {\color{blue}\bf --Yuan\vskip 2mm}}\vskip 2mm}
\def\lincomment#1{\vskip 2mm\boxit{\vskip 2mm{\color{blue}\bf#1} {\color{black}\bf --Lin\vskip 2mm}}\vskip 2mm}
\def\squarebox#1{\hbox to #1{\hfill\vbox to #1{\vfill}}}
\def\boxit#1{\vbox{\hrule\hbox{\vrule\kern6pt
          \vbox{\kern6pt#1\kern6pt}\kern6pt\vrule}\hrule}}

\def\theequation{\thesection.\arabic{equation}}
\newcommand{\ds}{\displaystyle}

\newcommand{\bJ}{\mbox{\bf J}}
\newcommand{\bF}{\mbox{\bf F}}
\newcommand{\bM}{\mbox{\bf M}}
\newcommand{\bR}{\mbox{\bf R}}
\newcommand{\bZ}{\mbox{\bf Z}}
\newcommand{\bX}{\mbox{\bf X}}
\newcommand{\bx}{\mbox{\bf x}}
\newcommand{\bww}{\mbox{\bf w}}
\newcommand{\bQ}{\mbox{\bf Q}}
\newcommand{\bH}{\mbox{\bf H}}
\newcommand{\bh}{\mbox{\bf h}}
\newcommand{\bz}{\mbox{\bf z}}
\newcommand{\br}{\mbox{\bf r}}
\newcommand{\ba}{\mbox{\bf a}}
\newcommand{\be}{\mbox{\bf e}}
\newcommand{\bG}{\mbox{\bf G}}
\newcommand{\bB}{\mbox{\bf B}}
\newcommand{\bb}{\mbox{\bf b}}
\newcommand{\bA}{\mbox{\bf A}}
\newcommand{\bC}{\mbox{\bf C}}
\newcommand{\bI}{\mbox{\bf I}}
\newcommand{\bD}{\mbox{\bf D}}
\newcommand{\bU}{\mbox{\bf U}}
\newcommand{\bc}{\mbox{\bf c}}
\newcommand{\bd}{\mbox{\bf d}}
\newcommand{\bs}{\mbox{\bf s}}
\newcommand{\bS}{\mbox{\bf S}}
\newcommand{\bV}{\mbox{\bf V}}
\newcommand{\bv}{\mbox{\bf v}}
\newcommand{\bW}{\mbox{\bf W}}
\newcommand{\bY}{\mathbf{ Y}}
\newcommand{\bw}{\mbox{\bf w}}
\newcommand{\bg}{\mbox{\bf g}}
\newcommand{\bu}{\mbox{\bf u}}
\newcommand{\mI}{\mbox{I}}

\def\bb{{\bf b}}

\newcommand{\bcU}{\boldsymbol{\cal U}}
\newcommand{\bbeta}{\boldsymbol{\beta}}
\newcommand{\bdelta}{\boldsymbol{\Delta}}
\newcommand{\bDelta}{\boldsymbol{\Delta}}
\newcommand{\boldeta}{\boldsymbol{\eta}}
\newcommand{\bxi}{\boldsymbol{\xi}}
\newcommand{\bGamma}{\boldsymbol{\Gamma}}
\newcommand{\bSigma}{\boldsymbol{\Sigma}}
\newcommand{\balpha}{\boldsymbol{\alpha}}
\newcommand{\bOmega}{\boldsymbol{\Omega}}
\newcommand{\btheta}{\boldsymbol{\theta}}
\newcommand{\bepsilon}{\boldsymbol{\epsilon}}
\newcommand{\bmu}{\boldsymbol{\mu}}
\newcommand{\bnu}{\boldsymbol{\nu}}
\newcommand{\bgamma}{\boldsymbol{\gamma}}
\newcommand{\btau}{\boldsymbol{\tau}}
\newcommand{\bTheta}{\boldsymbol{\Theta}}

\newtheorem{thm}{Theorem}
\newtheorem{lem}{Lemma}[section]
\newtheorem{rem}{Remark}[section]
\newtheorem{cor}{Corollary}[section]
\newcolumntype{L}[1]{>{\raggedright\let\newline\\\arraybackslash\hspace{0pt}}m{#1}}
\newcolumntype{C}[1]{>{\centering\let\newline\\\arraybackslash\hspace{0pt}}m{#1}}
\newcolumntype{R}[1]{>{\raggedleft\let\newline\\\arraybackslash\hspace{0pt}}m{#1}}

\newcommand{\tabincell}[2]{\begin{tabular}{@{}#1@{}}#2\end{tabular}}

\newcommand{\RN}[1]{%
  \textup{\uppercase\expandafter{\romannumeral#1}}%
}

\newcommand{\lline}[1]{\hline\multicolumn{#1}{c}{}\\[-1.3em]\hline}

\baselineskip=24pt
\begin{center}
{\Large \bf Dual-criterion  Dose Finding Designs Based on Dose-Limiting Toxicity and Tolerability}
\end{center}

\vspace{2mm}
\begin{center}
{\bf Yunlong Yang$^{1}$, Ying Yuan$^{1,*}$}
\end{center}

\noindent$^{1}$Department of Biostatistics, The University of Texas MD Anderson Cancer Center, Houston, TX, USA\\
$^*$Author for correspondence: yyuan@mdanderson.org 
\vspace{2mm}

\noindent \emph{\textbf{Abstract}}:  The primary objective of Phase I oncology trials is to assess the safety and tolerability of novel therapeutics. Conventional dose escalation methods identify the maximum tolerated dose (MTD) based on dose-limiting toxicity (DLT). However, as cancer therapies have evolved from chemotherapy to targeted therapies, these traditional methods have become problematic. Many targeted therapies rarely produce DLT and are administered over multiple cycles, potentially resulting in the accumulation of lower-grade toxicities, which can lead to intolerance, such as dose reduction or interruption. To address this issue, we proposed dual-criterion designs that find the MTD based on both DLT and non-DLT-caused intolerance. We considered the model-based design and model-assisted design that allow real-time decision-making in the presence of pending data due to long event assessment windows. Compared to DLT-based methods, our approaches exhibit superior operating characteristics when intolerance is the primary driver for determining the MTD and comparable operating characteristics when DLT is the primary driver.
\vspace{0.5cm}


\newpage
\section{Introduction}
\label{introduction}

Conventionally, the primary objective of phase I oncology trials has been to identify the maximum tolerated dose (MTD) based on dose-limiting toxicities (DLTs). These toxicities are typically defined as treatment-related grade 3 or higher non-hematologic toxicity or grade 4 or higher hematologic toxicity, according to the NCI Common Terminology Criteria for Adverse Events \citep{CTCAE2017}. For conventional cytotoxic compounds, which often induce severe and rapid-onset toxicities, DLTs occurring in the first cycle of treatment have served as reliable indicators of safety and tolerability \citep{Siu2009, Lin2022}. However, this framework presents challenges for molecularly targeted agents, immunotherapy, and antibody-drug conjugates. These novel therapies typically result in fewer DLTs compared to cytotoxic compounds, and patients often undergo longer treatment durations involving multiple cycles to maximize efficacy \citep{FDA2017}. Although patients may not experience DLTs, they may accumulate mild but disruptive toxicities over multiple treatment cycles, significantly impacting their quality of life. This often leads to dose reductions, interruptions, or discontinuations, which can compromise the actual efficacy of the treatment \citep{Paoletti2011, Theoret2022}. The U.S. Food and Drug Administration (FDA) has emphasized tolerability as a key consideration for dose optimization in Project Optimus. 

A number of approaches have been proposed to incorporate non-DLT toxicities in identifying the MTD. \citet{Thall2004} introduced the total toxicity burden, a weighted sum of toxicity grades across different types, for dose finding. \citet{Yuan2007} presented the equivalent toxicity score, converting various toxicity grades into an equivalent number of DLTs, and developed the quasi-CRM (continuous reassessment method) design. \citet{Lee2011, Lee2012} proposed the continual reassessment method with multiple toxicity constraints using an ordinal probit model, and utilized regression on historical data to derive severity weights, defining the toxicity burden score. \citet{Ezzalfani2013} suggested the total toxicity profile, defined as the Euclidean norm of severity weights of toxicities, to summarize overall severity. \citet{Mu2018} introduced the generalized Bayesian optimal interval design (gBOIN), offering a unified framework to accommodate various toxicity grade and type scoring systems for dose finding. \citet{Lin2018} developed the model-assisted multiple-toxicity Bayesian optimal interval (MT-BOIN) design to determine the MTD satisfying multiple toxicity constraints, with a decision table for ease of implementation.  \citet{Jiang2024} and \citet{Chen2024} extended the keyboard design to accommodate low grade toxicities.


A practical challenge with existing methods is the multitude and heterogeneity of different types and grades of toxicities. Phase I trials commonly involve the assessment of up to 20-30 different types of toxicities. Moreover, toxicities such as low-grade diarrhea can have significantly different clinical implications depending on their duration. Prolonged occurrences cause much greater patient burden than brief ones \citep{FDA2017}. Consequently, assigning appropriate severity scores to each type of grade 1 and 2 toxicity proves extremely challenging, given the large variation and subjectivity involved. This limitation restricts the adoption of the aforementioned designs in practice.


To tackle this challenge, we propose using tolerability as an aggregated endpoint to quantify the consequence and impact of low-grade toxicities, rather than modeling different types and grades of toxicities. Here, tolerability is defined as a binary endpoint indicating whether dose reduction, interruption, or discontinuation occurs due to non-DLT toxicities. This simplification transforms complex multidimensional toxicity endpoints into a single endpoint that directly reflects the clinical impact of toxicities. Additionally, the tolerability endpoint is more measurable and requires less subjective input from physicians \citep{McKee2016},  compared to specifying severity scores for different types, grades, and durations of toxicities.

One challenge of using tolerability as the endpoint is the time it takes (multiple treatment cycles) to ascertain. We address this issue using Bayesian data augmentation, which predicts the pending tolerability endpoint based on observed data. We consider both a model-based approach and a model-assisted approach that incorporate both DLT and non-DLT-caused intolerance to determine the MTD. We refer to our designs as dual-criterion designs to emphasize their use of both DLT and non-DLT-caused intolerance in dose escalation and selection. Simulations demonstrate that the proposed designs outperform the approach that only considers DLT.

It's worth noting that under Project Optimus, which emphasizes the consideration of risk-benefit assessment of doses, identifying the MTD remains important as it establishes the range of safe doses, forming the basis for further optimization (e.g., randomizing patients into MTD and a lower dose to further compare safety, tolerability, efficacy, PK and PK). \citet{Yuan2024} provides a comprehensive review of approaches for dose optimization and refer to it as a two-stage strategy.

The remainder of this article is organized as follows: Section 2 introduces the models and decision rules of the proposed model-based and model-assisted dual-criterion designs. Section 3 presents simulation results comparing the proposed methods with conventional approaches, as well as sensitivity analysis on the proposed methods. Section 4 concludes with a discussion.

\section{Method}\label{method}
In a phase I trial with $J$ doses, $d_1< \cdots <d_J$,  under investigation, let $Y_T$ and $Y_R$ denote indicator variables for the occurrence of DLT and intolerance due to non-DLT toxicities, respectively. $Y_R$ is intended to capture intolerance due to low-grade non-DLT toxicities. If a patient cannot tolerate a dose due to DLTs, $Y_T=1$ and $Y_R=0$. We assume that the assessment window for $Y_T$ and $Y_R$ are $T_T$ and $T_R$, respectively. In practice, $T_T$ is often the first treatment cycle, whereas $T_R$ spans multiple treatment cycles, e.g., 3 cycles. For example, with a treatment cycle of 21 days, the DLT window $T_T=21$ days and tolerance window $T_R = 63$ days. The values of $T_T$ and $T_R$ should be chosen based on trial characteristics and be sufficiently long to capture potential DLT and tolerability issues. As described previously, a key challenge is that $T_R$ is relatively long, thus $Y_R$ may not be observed at the time of making dose assignment decisions for the next cohort, posing a major challenge for conducting the trial. In what follows, we first describe a model-based time-to-event dual-criterion design (referred to as TITE-DC), followed by a model-assisted time-to-event dual-criterion design (referred to as TITE-BOIN\textsubscript{DC}), to address this design challenge. Our designs are also able to handle the case where $Y_T$ is not quickly observed for decision-making, e.g., DLT is also assessed over multiple cycles.


\subsection{Model-based TITE-DC Design} \label{DC}
\subsubsection{Statistical Model }
We model the joint distribution of $(Y_T, Y_R)$ using a latent variable approach. Let $Z_T$ and $Z_R$ denote normally distributed latent variables, which are related to $(Y_T, Y_R)$ as follows:
\begin{eqnarray}
    Y_{k} = 
    \begin{cases}
      0, & \text{if}\ Z_{k} < 0 \\
         1, & \text{if}\ Z_{k} \ge 0
    \end{cases} ;\quad k\in(T,R).
\end{eqnarray}
$Z_T$ and $Z_R$ can be interpreted as the patient's
latent traits, and $Y_T$ and $Y_R$ are the clinical manifestations of unobserved $Z_T$ and $Z_R$. When
$Z_T$ and $Z_R$  pass the threshold of zero, certain clinical outcomes (e.g., DLT, intolerance) are
observed.

At a given dose, we model $(Z_T, Z_R)$ using a bivariate normal regression model:
\begin{eqnarray}\label{bivnormal}
    \begin{pmatrix}
        Z_{T} \\ Z_{R} 
    \end{pmatrix}
        \sim N_2\left(
    \begin{pmatrix}
        \alpha_T+\beta_Td^*_j \\ \alpha_R+\beta_Rd^*_j
    \end{pmatrix},
    \begin{pmatrix} 
        \sigma^2_T & \rho\\ \rho & \sigma^2_R 
    \end{pmatrix}\right),
\end{eqnarray}
where $\alpha$ and $\beta$ are regression parameters, $d^*_j={d_j}/{d_J}$ is the standardized dose at level $j$, and $\rho$ quantifies the correlation between $Z_T$ and $Z_R$ (or equivalently $Y_T$ and $Y_R$). Following \citet{AlbertChib1993}, we fix variance $\sigma^2_k=1$ to make the model identifiable. 

As described previously, the key challenge is that $Y_R$ is typically not observed at the time of making dose assignment decisions for the next cohort. Under the Bayesian paradigm, one intuitive way to address this missing data issue is to use Bayesian data augmentation, under which we draw the unobserved $Y_R$ from its posterior predictive distribution. This approach has been previously used by \citet{Liu2014} in their data-augmentation continuous reassessment method (DA-CRM), as well as by \citet{Jin2014}, \citet{Zhou2022}, and \citet{Takeda2023} to address delayed toxicity and efficacy endpoints in phase I/II trial designs.

To proceed, let $i$ index subjects, and $X_{Ti}$ and $X_{Ri}$ represent the time to DLT and intolerance, respectively. For patients who have no DLT nor intolerance issues, $X_{Ti}$ and $X_{Ri}$ can be set as large fixed values greater than $T_T$ and $T_R$, respectively. Let $t_i$ denote the follow-up time for the $i$th enrolled patient at the time of making dose assignment decisions for the next cohort. Following the work of \citet{Zhou2022}, we make the following assumptions about $X_{Ti}$ and $X_{Ri}$. We suppress $i$ for brevity:
\begin{enumerate}
    \item $X_{T}$ and $X_{R}$ are conditionally independent: $\Pr(X_T>t,X_R>t \,|\,Y_T=1,Y_R=1)=\Pr(X_T>t \,|\, Y_T=1)\Pr(X_R>t \,|\, Y_R=1)$. This is a reasonable assumption based on how we defined intolerance, and because $T_T$ is typically shorter than $T_R$ for targeted therapy trials.
    \item $X_{T}$ and $X_{R}$ are uniformly distributed over their respective assessment windows (e.g., for time to DLT, $\Pr(X_T>t \,|\, Y_T=1)=1- {t}/{T_T}$). This assumption has been shown to yield robust performance and is widely used in early phase trial designs \citep{Cheung2000, Liu2014, Zhou2022, Takeda2023}.
    \item $X_{T}$ and $X_{R}$ are independent from $d_j$ conditional on $Y_T=1$ and $Y_R=1$. This assumption has been routinely applied in previous designs involving late-onset toxicity  \citep{Cheung2000, Liu2014, Zhou2022, Takeda2023}.
\end{enumerate}
By imposing structural constraints, these assumptions allow more efficient use of the (very limited) observed data from small phase I trials, thereby stabilizing the estimation and decision making. In addition, they also simplify  the imputation of unobserved $Y_{Ri}$ and $Y_{Ti}$. 

At the time of decision-making, there are three possible missing data patterns: i) both $Y_{Ti}$ and $Y_{Ri}$ are pending, ii) $Y_{Ti}$ is observed but $Y_{Ri}$ is pending, and iii) $Y_{Ti}$ is pending but $Y_{Ri}$ is observed. Let $p_{ab} = \Pr(Y_T=a,Y_R=b)$, where $a, b \in (0, 1)$. For these three patterns, the pending values can be drawn from their posterior predictive distributions as follows (see Supplementary Materials for the derivation):
\begin{itemize}
\item For patients with $Y_{T}$ and $Y_{R}$ pending, we draw the unobserved ($Y_{T}, Y_{R}$) from the following posterior distribution (suppressing $i$ for brevity):
\vspace*{-\abovedisplayskip} 
\begin{flalign*}
& \Pr(Y_T=1,Y_R=1 \,|\, X_T>t,X_R>t) && \\ 
&\quad = \frac{(1-\frac{t}{T_T})(1-\frac{t}{T_R})p_{11}}{p_{00}+(1-\frac{t}{T_R})p_{01}+(1-\frac{t}{T_T})p_{10}+(1-\frac{t}{T_T})(1-\frac{t}{T_R})p_{11}}
\end{flalign*}
\begin{flalign*}
& \Pr(Y_T=1,Y_R=0 \,|\, X_T>t,X_R>t) && \\ 
&\quad = \frac{(1-\frac{t}{T_T})p_{10}}{p_{00}+(1-\frac{t}{T_R})p_{01}+(1-\frac{t}{T_T})p_{10}+(1-\frac{t}{T_T})(1-\frac{t}{T_R})p_{11}} 
\end{flalign*}
\begin{flalign*}
& \Pr(Y_T=0,Y_R=1 \,|\, X_T>t,X_R>t) && \\ 
&\quad = \frac{(1-\frac{t}{T_R})p_{01}}{p_{00}+(1-\frac{t}{T_R})p_{01}+(1-\frac{t}{T_T})p_{10}+(1-\frac{t}{T_T})(1-\frac{t}{T_R})p_{11}} 
\end{flalign*}
\begin{flalign*}
& \Pr(Y_T=0,Y_R=0 \,|\, X_T>t,X_R>t) && \\ 
&\quad = \frac{p_{00}}{p_{00}+(1-\frac{t}{T_R})p_{01}+(1-\frac{t}{T_T})p_{10}+(1-\frac{t}{T_T})(1-\frac{t}{T_R})p_{11}}.
\end{flalign*}

\item For patients with $Y_{T}$ observed and $Y_{R}$ pending, we draw the unobserved $Y_{R}$ from the following posterior distribution:
\vspace*{-0.5em} 
\begin{flalign*}
 \Pr(Y_T =y_T, Y_R=1 \,|\, X_R>t) 
 = \left[\frac{(1-\frac{t}{T_R})p_{11}}{(1-\frac{t}{T_R})p_{11}+p_{10}} \right]^{y_T} 
\left[\frac{(1-\frac{t}{T_R})p_{01}}{(1-\frac{t}{T_R})p_{01}+p_{00}} \right]^{1-y_T}.
\end{flalign*}

\item For patients with $Y_{T}$ pending and $Y_{R}$ observed, we draw the unobserved $Y_{T}$ from the following posterior distribution:
\vspace*{-\abovedisplayskip} 
\begin{flalign*}
 \Pr(Y_T=1,Y_R=y_R \,|\, X_T>t)  = \left[\frac{(1-\frac{t}{T_T})p_{11}}{(1-\frac{t}{T_T})p_{11}+p_{01}} \right]^{y_R} 
\left[\frac{(1-\frac{t}{T_T})p_{10}}{(1-\frac{t}{T_T})p_{10}+p_{00}} \right]^{1-y_R}.
\end{flalign*} 

\end{itemize} 
To fit the model, we perform Bayesian data augmentation as follows:
\begin{enumerate}
    \item Initialize $p_{ab}$ with their posterior estimates, obtained by fitting the proposed model (\ref{bivnormal}) to the observed data using Gibbs sampling, ignoring pending $Y_T$ and $Y_R$.
    \item  I (imputation) step: draw pending $Y_{T}$ and/or $Y_{R}$ according to the patient's missing data pattern, as described above, to acquire a complete dataset. 
    \item P (posterior) step: draw the unknown parameters based on the complete dataset, using Gibbs sampling, and update $p_{ab}$.  
    \item Iterate between the I and P steps until the algorithm converges.
\end{enumerate}


\subsubsection{Prior Specification}
A hallmark of phase I trials is small sample sizes. Thus, it is critical to appropriately specify the prior: it should be sufficiently vague such that data will dominate the posterior and drive decision-making, while also not being too diffuse (e.g., with a very large variance). Diffuse priors can cause estimation instability due to limited data and a high likelihood of drawing extreme parameter values, especially at the early stage of the trial when only a few cohorts of patients are treated. We follow the approach of  \citet{Guo2017} to specify the priors. 

Defining $\pi_{kj}=p(Y_k=1 \,|\, d_j)$ and integrating out the latent variables $Z_k$ in model (\ref{bivnormal}), $\pi_{kj}$ follows a probit model:  ${\rm probit}(\pi_{kj})=\alpha_k+\beta_kd^*_j$, where $k=T, R$.
Therefore, we assign a normal $N(0,1.25^2)$ prior to the intercept $\alpha_k$. This prior spans the probability between 0.006 and 0.994 when $\alpha_k$ deviates from its prior mean by one standard deviation on each side. We assign the slope $\beta_k$ a half-normal $N(0,1.24^2)$ prior to ensure that toxicity monotonically increases with dose. This prior is centered at 1 to reduce positive bias due to truncation when the true value is close to 0. We set the standard deviation of 1.24 such that a standard deviation change will move the probit probability to 0.994 at $d_J$ and $\alpha_k=0$. Finally, we assign a positive uniform prior $U(0,1)$ to the correlation parameter $\rho$, as we expect DLT to be positively correlated with intolerance.


\subsubsection{Dose-finding algorithm} \label{DCFA}
Let $\phi_T$ and $\phi_R$ represent pre-specified target probabilities for $Y_T$ and $Y_R$, respectively. 
The dose-finding algorithm of TITE-DC is described as follows:
\begin{enumerate}
\item Treat the first cohort of patients at $d_1$ or a prespecified starting dose.
\item Based on observed interim data $D$, obtain the posterior estimates of ${\pi}_{kj}$ (denoted as $\hat{\pi}_{kj}$), and identify the dose with
$\hat{\pi}_{Tj}$ being closest to $\phi_T$ and the dose with  $\hat{\pi}_{Rj}$ being closest to $\phi_R$. Let $j^*$ denote the lower dose level between the two identified doses, defined as $ j^*=\min(\argmin_{j\in(1,...,J)}|\hat{\pi}_{kj}-\phi_k|)$ where $k=T, R$. Let $j^c$ denote the current dose level. Perform dose escalation or de-escalation as follows:
\begin{itemize}
    \item If $j^c<j^*$ and $j^c<J$, escalate the dose level to $j^c+1$.
    \item If $j^c>j^*$ and $j^c>1$, de-escalate the dose level to $j^c-1$.
    \item Otherwise, retain the dose level at $j^c$.
\end{itemize}
\item Repeat Step 2 until reaching the maxium sample size. Based on the final data, identify $j^*$ and select it as the MTD. 
\end{enumerate} 

In Step 2, the dose escalation/de-escalation decision is based on both  $Y_T$ and $Y_R$ (i.e., taking the more conservative one), which distinguish it from conventional dose-finding designs where the dose decision depends solely on $Y_T$. 

To safeguard patients from overly toxic or intolerable doses, we impose the following overdosing and early stopping rule: during the trial, if $\Pr(\pi_{Tj} > \phi_T \,|\, D)>0.95$ and/or $\Pr(\pi_{Rj} > \phi_R \,|\, D)>0.95$, eliminate dose level $j$ and all higher dose levels. If all dose levels are eliminated, the trial should be terminated, and no dose should be selected as the MTD. Additionally, to reduce the uncertainty caused by pending data, patient enrollment and dose decisions will be suspended when the ratio of patients with pending to non-pending DLT and intolerance data at the current dose is $\ge 0.5$. 

\subsection{Model-Assisted TITE-BOIN\textsubscript{DC} Design} \label{BOINDC}
\subsubsection{Statistical Model}
In this section, we introduce a model-assisted dual-criterion design, TITE-BOIN\textsubscript{DC}, as an extension of the TITE-BOIN design \citep{Yuan2018}. Unlike TITE-DC, which models dose-toxicity and dose-intolerance curves, TITE-BOIN\textsubscript{DC} directly estimates $\pi_{Tj}$ and $\pi_{Rj}$ using the sample mean based on the local data observed at dose level $j$, making it simpler to implement and also more robust.

Let $n_j$ denote the number of patients dosed at level $j$, and $m_{Tj}$ and $m_{Rj}$ denote the observed number of patients experiencing DLT and intolerance. When there is no pending $Y_T$ and $Y_R$, $\pi_{Tj}$ and $\pi_{Rj}$ are simply estimated by their sample means $\hat{\pi}_{Tj} = m_{Tj}/n_j$ and $\hat{\pi}_{Rj} = m_{Rj}/n_j$. As described previously, the challenge is that often some $Y_T$ and $Y_R$ are pending at decision times. 

We first consider the estimation of  $\pi_{Rj}$ when some of $Y_{R}$ is pending. Following the approach of TITE-BOIN, we first perform single imputation by replacing the pending $Y_R$ with its expectation. Given the imputed complete data, we then estimate $\hat{\pi}_{Rj}$ using its sample mean. Let $M_R$ represent the set of patients with pending $Y_R$. For $i\in M_R$, the corresponding pending $Y_R$ can be imputed by
\begin{equation} \label{imputeR}
\hat{Y}_{Ri} = \Pr(Y_{Ri} = 1 \,|\, X_{Ri} > t_i, d_j) = \frac{\pi_{Rj}(1-\frac{t_i}{T_R})}{\pi_{Rj}(1-\frac{t_i}{T_R})+(1-\pi_{Rj})}{n_j},
\end{equation}
where unknown probability $\pi_{Rj}$ can be replaced by its posterior mean estimate based on the observed data, using a beta-binomial model $m_{Rj} \sim Binomial(n_j, \pi_{Rj})$ with a vague prior $\pi_{Rj}\sim Beta(\alpha_R, \beta_R)$, $\alpha_R=0.5\phi_R$ and $\beta_R=1-\alpha_R$, representing an effective sample size of $1$. As noted by \citet{Yuan2018}, this observed-data-based estimate slightly overestimates $\pi_{Rj}$, but this is not a concern as it leads to a conservative estimate of $Y_{Ri}$, thereby improving the safety of the design. Given the imputed $\hat{Y}_{Ri}$, an estimate of $\pi_{Rj}$ is given by
$$ \hat{\pi}_{Rj} = \frac{m_{Rj} + \sum_{i\in M_{Rj}}\hat{Y}_{Ri} }{n_j}.$$

The estimation of  $\pi_{Tj}$, when some of $Y_{T}$ is pending, can be performed in a similar way as above, by imputing pending $Y_{Ti}$ as
\begin{equation} \label{imputeT}
\hat{Y}_{Ti}  = \frac{\pi_{Tj}(1-\frac{t_i}{T_T})}{\pi_{Tj}(1-\frac{t_i}{T_T})+(1-\pi_{Tj})} {n_j}  \approx \frac{\pi_{Tj}(1-\frac{t_i}{T_T})}{(1-\pi_{Tj})}.
\end{equation}
The approximation on the right-hand side of the equation is appropriate as $\pi_{Tj}$ is often small in most phase 1 trials. This approximation is optional. It simplifies the calculation and is used by TITE-BOIN. Similarly, the unknown probability $\pi_{Tj}$ can be replaced by its posterior mean estimate based on observed data, using a beta-binomial model.

\subsubsection{Dose-finding algorithm} 
Let $\lambda_{Te}$ and $\lambda_{Td}$ denote the BOIN dose escalation and de-escalation boundaries, respectively, for the target $\phi_T$. Define $\lambda_{Re}$ and $\lambda_{Rd}$ similarly for the target $\phi_R$.
The dose-finding algorithm of TITE-BOIN\textsubscript{DC} is described as follows:
\begin{enumerate}
\item Treat the first cohort of patients at $d_1$ or a prespecified starting dose.
\item Based on observed interim data $D$, calculate $\hat{\pi}_{Rj}$ using (\ref{imputeR}) and $\hat{\pi}_{Tj}$ using (\ref{imputeT}), and apply their respective BOIN boundaries to determine their individual recommended doses $j^*_k$ for $Y_k$, $k=T, R$, as follows:
\begin{itemize}
    \item If $\hat{\pi}_{kj} \leq \lambda_{ke}$ and $j^c<J$, set $j^*_k = j^c+1$.
    \item If $\hat{\pi}_{kj} \geq \lambda_{kd}$ and $j^c>1$,  set $j^*_k = j^c- 1$.
    \item Otherwise, set $j^*_k = j^c$.
\end{itemize}
Perform dose escalation/de-escalation by treating the next cohort of patients at dose level $j^*= \min (j^*_T, \, j^*_R)$.
\item Repeat Step 2 until reaching the maxium sample size. Based on the final data, select the MTD as $j^*= \min(\argmin_{j\in(1,...,J)}|\hat{\pi}_{kj}-\phi_k|)$, $k=T, R$, where $\hat{\pi}_{Tj}$ and $\hat{\pi}_{Rj}$ are isotonic estimates of DLT and intolerance rate, based on isotonic regression \citep{Yuan2016}.
\end{enumerate} 

The same overdosing and early stopping rule as described in Section \ref{DCFA} will be used, except that $\Pr(\pi_{kj} > \phi_k \,|\, D)$ will be evaluated based on  a beta-binomial model: $m_{kj} \sim Binomial(n_j, \pi_{kj})$ with a vague prior $\pi_{kj}\sim Beta(1, 1)$. Additionally, the same accrual suspension as in Section \ref{DCFA} will be employed to reduce the uncertainty resulting from excessive pending data.

\section{Simulation}
\subsection{Simulation Setting}

We performed simulation to evaluate the operating characteristics of TITE-DC and TITE-BOIN\textsubscript{DC} designs. 
We considered $J=5$ dose levels with target $\phi_T = 0.25$ for DLT, and $\phi_R = 0.5$ for intolerance. Patients are enrolled in cohorts of 3, with the first cohort receiving the lowest dose level. The maximum sample size is $N=30$ patients, consisting of 10 cohorts. The treatment cycle is 21 days. The DLT assessment window is the first cycle, and the intolerance assessment window is 3 cycles. The enrollment follows a Poisson process with a rate of 1 patient per 10 days. Among patients experiencing DLT and/or intolerance, the times to event $X_T$ and  $X_R$ are generated from uniform distributions.

We constructed 11 scenarios to cover different shapes of dose-DLT and dose-intolerance curves (Table 1). Scenarios 1-7 consider cases where the target dose is driven by intolerance rather than DLT. Scenarios 8-9 consider cases where the target dose is driven by DLT. Scenarios 10-11 consider cases where the target dose is driven by both DLT and intolerance. 

We compared TITE-DC and TITE-BOIN\textsubscript{DC} with their complete-data versions (denoted as DC and BOIN\textsubscript{DC}, respectively), which require $Y_T$ and $Y_R$ to be fully observed before enrolling the next cohort of patients. These complete-data versions provide a useful benchmark to evaluate how well our methods handle pending data, although they are often infeasible in practice due to long trial durations caused by frequent enrollment suspension. To demonstrate the importance of considering both DLT and intolerance, we also compared TITE-DC and TITE-BOIN\textsubscript{DC} against the BOIN design, which considers only DLT. Previous research shows that CRM has similar operating characteristics as BOIN \citep{Zhou2018}, so we included only BOIN for comparison. In each scenario, we simulated 1000 trials.


\subsection{Operating Characteristics}

Scenarios 1-7 consider cases where the MTD is driven by intolerance rather than DLT. In scenario 1 for example, the MTD is dose level 3, for which the intolerance rate equals the target $\phi_R = 0.5$, but the DLT rate is lower than the target $\phi_T = 0.25$. If only DLT is considered, ignoring intolerance, the MTD would be dose level 5. In this case, the percentage of correct selection (PCS) of the MTD is 65.8\% for TITE-DC and 61.9\% for TITE-BOIN\textsubscript{DC}, substantially higher than the 24.6\% PCS for BOIN. Since BOIN only considers DLT, it tends to select dose level 5 as the MTD, which has an unacceptable intolerance rate of 0.9. This highlights the importance of considering both DLT and intolerance.

The performance of TITE-DC and TITE-BOIN\textsubscript{DC} is comparable to their complete-data counterparts, DC and BOIN\textsubscript{DC}, showing that our methods handle pending data efficiently. DC and BOIN\textsubscript{DC} require repeated suspensions, leading to substantially longer trial durations (e.g., 28.7 and 29.0 months) compared to TITE-DC and TITE-BOIN\textsubscript{DC} (18.3 and 17.6 months). This demonstrates that TITE-DC and TITE-BOIN\textsubscript{DC} can dramatically shorten trial duration without compromising the performance in identifying the MTD. Similar results are observed in scenarios 2-7. 

Between TITE-DC and TITE-BOIN\textsubscript{DC}, neither dominates the other in terms of PCS. In the scenarios where the target is a higher dose (e.g., scenarios 1, 2), TITE-DC has a higher PCS, while in the scenarios where the target is a lower dose (e.g., scenarios 3,4), TITE-BOIN\textsubscript{DC} achieves a higher PCS. In terms of safety, TITE-BOIN\textsubscript{DC} outperforms TITE-DC in 6 out of 7 scenarios, with a lower percentage of overdosed patients. For example, in scenario 2, TITE-DC treated 11.9\% of patients at doses above the MTD, whereas TITE-BOIN\textsubscript{DC} treated only 4.3\% of patients at doses above the MTD.

Scenarios 8-9 consider cases where the target dose is driven by DLT, which favors the DLT-based BOIN design. The PCS of TITE-DC and TITE-BOIN\textsubscript{DC} remain competitive. Specifically, the PCS of TITE-DC is slightly lower than that of BOIN, while the PCS of TITE-BOIN\textsubscript{DC} is comparable to that of BOIN. The PCS of TITE-DC and TITE-BOIN\textsubscript{DC} are similar to DC and BOIN\textsubscript{DC}, demonstrating high efficiency in handling pending data. Between TITE-DC and TITE-BOIN\textsubscript{DC}, TITE-BOIN\textsubscript{DC} appears safer with a higher PCS.

Scenarios 10-11 consider cases where the MTD is driven by both DLT and intolerance. In these scenarios, TITE-DC and TITE-BOIN\textsubscript{DC} both achieve higher PCS compared to BOIN. This demonstrates an additional benefit of utilizing dual-criterion when both DLT and intolerance indicate the same dose as the  MTD. 

\subsection{Sensitivity Analysis}
One critical assumption used by TITE-DC and TITE-BOIN\textsubscript{DC} to handle pending $Y_T$ and $Y_R$ is that $X_T$ and $X_R$ are uniformly distributed over $(0, T_T)$ and $(0, T_R)$, respectively. We evaluate the sensitivity of TITE-DC and TITE-BOIN\textsubscript{DC} to this assumption by simulating $X_R$ from a piecewise uniform distribution, where each treatment cycle has different probabilities of intolerance. Within each cycle, the probability of intolerance is uniformly distributed. Let $(w_1, w_2, w_3)$ denote probabilities of intolerance in cycles 1 to 3. We consider four different piecewise uniform distributions for $X_R$ with $(w_1, w_2, w_3)$ = (1/3, 1/3, 1/3), (1/7, 2/7, 4/7), (1/10, 1/10, 8/10), (8/10, 1/10, 1/10). The first distribution corresponds to a uniform distribution over the assessment window; the second distribution places higher weights towards later cycles; and the third and fourth distributions place most of the weight in cycles 3 and 1, respectively. We selected scenario 2 from Table 1 to evaluate model sensitivity. Figure 1 depicts the operating characteristics of TITE-DC and TITE-BOIN\textsubscript{DC}, demonstrating the robustness of TITE-DC and TITE-BOIN\textsubscript{DC} to the violation of the uniform time-to-event assumption. The PCS, the number of patients allocated to the MTD, and trial duration are comparable across the four distributions. The percentage of overdoses is higher in the second and third distributions due to more delayed events in these two cases compared to the first and fourth distributions.

We also evaluated the sensitivity of TITE-DC and TITE-BOIN\textsubscript{DC} to the accrual rate. As noted by \citet{Yuan2018}, the percentage of pending patients is determined by the ratio of the accrual rate to the assessment window. There is a one-to-one correspondence between varying the accrual rate with a fixed assessment window and varying the assessment window with a fixed accrual rate. Therefore, our simulation results are applicable to varying assessment windows. Figure 2 depicts the operating characteristics of TITE-DC and TITE-BOIN\textsubscript{DC} under varying accrual rates, which are generally similar and not sensitive to the accrual rate except the trial duration. The trial duration is expected to increase with slower accrual.

\section{Discussion}
\label{discussion}
We have proposed dual-criterion dose-finding designs that determine the MTD based on both DLT and intolerance caused by non-DLT events. We considered both the model-based TITE-DC design and the model-assisted TITE-BOIN\textsubscript{DC} design. Simulation studies show that both designs outperform DLT-based dose-finding designs and can efficiently handle pending data caused by the potentially long assessment window for DLT and intolerance. TITE-BOIN\textsubscript{DC} is generally safer than TITE-DC. Consistent with this, TITE-BOIN\textsubscript{DC} has a higher probability of identifying the MTD when it is located at a lower dose, whereas TITE-DC has a higher probability of identifying the MTD when it is located at a higher dose.

In this paper, we focus on determining the MTD. As the paradigm of dose-finding shifts from the identification of the MTD to the optimal biological dose (OBD), different extensions can be considered. One approach takes a two-stage design strategy: after establishing the MTD based on TITE-DC or TITE-BOIN\textsubscript{DC} designs, we randomize patients into the MTD and a dose lower than the MTD to collect additional data to determine the OBD. The other approach is the efficacy-integrated strategy \citep{Yuan2024}, where we simultaneously consider efficacy, toxicity, and tolerability to guide dose finding. For the model-based approach, we can introduce an efficacy endpoint and accommodate it using a trivariate normal latent-variable model to account for efficacy, DLT, and intolerance. For the model-assisted design, we can apply BOIN boundaries to the efficacy endpoint and incorporate it into the dose-finding algorithm, or model the risk-benefit tradeoff as in the BOIN12 design \citep{BOIN12}.

When treatment is administered over multiple cycles, the likelihood that patients drop out of the study increases. For example, patients may be off the study due to disease progression, resulting in missing data. Incorporating the missing data into our design is a topic of interest. Methods have been developed to incorporate partial information in the context of patient dropout \citep{Guo2015}, which may be adapted to our designs.

\newpage


\begin{table}[]
\centering
\caption{Operating Characteristics of TITE-DC, TITE-BOIN\textsubscript{DC}, their non-TITE counterparts, and BOIN, under 11 scenarios. The MTDs are bolded.}
\label{tb: TITE-DC1}
\resizebox{0.95\textwidth}{!}{%
\begin{tabular}{lllcccccccc} \hline \hline
 &  &  & \multicolumn{5}{c}{\textbf{Dose}} & \textbf{Duration} & \textbf{Overdose} \\
  \cmidrule{4-8}
 & \textbf{Design}  &  & $\bm{d_1}$ & $\bm{d_2}$ & $\bm{d_3}$ & $\bm{d_4}$ & $\bm{d_5}$ & \textbf{(Months)} &  \textbf{(\%)} \\ \hline
  &  &  & \multicolumn{5}{c}{\textbf{Scenario 1}} \\ 
 &  & DLT & 0.05 & 0.10 & 0.15 & 0.20 & \textbf{0.25} \\
 &  & Intolerance & 0.10 & 0.30 & \textbf{0.50} & 0.70 & 0.90 \\ \hline
 & TITE-DC & Selection  $\%$ & 0.8 & 16.7 & \textbf{65.8} & 16.2 & 0.5 \\
 &  & No. patients & 5.8 & 8.5 & \textbf{10.9} & 4.3 & 0.5 & 16.9 & 15.9 \\
 & TITE-BOIN\textsubscript{DC} & Selection  $\%$ & 0.8 & 29.4 & \textbf{61.9} & 8.0 & 0.1 \\
 &  & No. patients & 6.3 & 11.7 & \textbf{9.4} & 2.4 & 0.3 & 16.4 & 8.8 \\
 & DC & Selection  $\%$ & 0.2 & 16.2 & \textbf{66.8} & 15.9 & 0.8 \\
 &  & No. patients & 5.4 & 8.3 & \textbf{11.8} & 4.1 & 0.4 & 28.3 & 14.9 \\
 & BOIN\textsubscript{DC} & Selection  $\%$ & 0.7 & 28.4 & \textbf{61.1} & 9.8 & 0 \\
 &  & No. patients & 5.9 & 11.3 & \textbf{10.2} & 2.4 & 0.2 & 28.5 & 8.8 \\
 & BOIN & Selection  $\%$ & 0.6 & 8.6 & \textbf{24.6} & 31.3 & 35.0 \\
 &  & No. patients & 5.0 & 6.8 & \textbf{7.4} & 6.0 & 4.9 & 16.0 & 36.3 \\

&  &  & \multicolumn{5}{c}{\textbf{Scenario 2}} \\
 &  & DLT & 0.05 & 0.10 & 0.15 & 0.20 & \textbf{0.25} \\
 &  & Intolerance & 0.05 & 0.10 & 0.30 & \textbf{0.50} & 0.70 \\ \hline
 & TITE-DC & Selection  $\%$ & 0.1 & 4.2 & 24.0 & \textbf{54.6} & 17.1 \\
 &  & No. patients & 5.3 & 5.6 & 7.6 & \textbf{8.0} & 3.6 & 18.3 & 11.9 \\
 & TITE-BOIN\textsubscript{DC} & Selection  $\%$ & 0.4 & 9.0 & 44.4 & \textbf{41.3} & 5.0 \\
 &  & No. patients & 5.3 & 8.0 & 9.8 & \textbf{5.7} & 1.3 & 17.6 & 4.3 \\
 & DC & Selection  $\%$ & 0.5 & 5.1 & 23.5 & \textbf{52.3} & 18.6 \\
 &  & No. patients & 5.1 & 5.5 & 7.4 & \textbf{8.3} & 3.7 & 28.7 & 12.2 \\
 & BOIN\textsubscript{DC} & Selection  $\%$ & 0.5 & 8.5 & 43.0 & \textbf{42.5} & 5.6 \\
 &  & No. patients & 5.1 & 7.7 & 9.8 & \textbf{6.1} & 1.3 & 29.0 & 4.3 \\
 & BOIN & Selection  $\%$ & 0.6 & 8.5 & 24.5 & \textbf{31.4} & 35.2 \\
 &  & No. patients & 5.0 & 6.7 & 7.3 & \textbf{6.0} & 4.9 & 16.0 & 16.4 \\

  &  &  & \multicolumn{5}{c}{\textbf{Scenario 3}} \\ 
 &  & DLT & 0.05 & 0.10 & 0.15 & 0.20 & \textbf{0.25} \\
 &  & Intolerance & 0.30 & \textbf{0.50} & 0.70 & 0.90 & 0.95 \\ \hline
 & TITE-DC & Selection  $\%$ & 14.1 & \textbf{66.6} & 17.1 & 0.2 & 0 \\
 &  & No. patients & 11.7 & \textbf{12.6} & 4.7 & 0.5 & 0 & 15.3 & 17.7 \\
 & TITE-BOIN\textsubscript{DC} & Selection  $\%$ & 19.5 & \textbf{70.6} & 9.9 & 0 & 0 \\
 &  & No. patients & 12.4 & \textbf{13.4} & 3.8 & 0.4 & 0 & 15.0 & 14.1 \\
 & DC & Selection  $\%$ & 15.1 & \textbf{67.4} & 15.6 & 0.4 & 0 \\
 &  & No. patients & 11.1 & \textbf{13.8} & 4.4 & 0.4 & 0 & 28.0 & 16.1 \\
 & BOIN\textsubscript{DC} & Selection  $\%$ & 19.6 & \textbf{69.1} & 11.0 & 0 & 0 \\
 &  & No. patients & 11.7 & \textbf{14.3} & 3.6 & 0.3 & 0 & 28.0 & 13.1 \\
 & BOIN & Selection  $\%$ & 0.6 & \textbf{8.5} & 24.5 & 31.3 & 35.2 \\
 &  & No. patients & 5.0 & \textbf{6.7} & 7.3 & 6.0 & 4.9 & 16.0 & 60.9 \\

&  &  & \multicolumn{5}{c}{\textbf{Scenario 4}} \\ 
 &  & DLT & 0.05 & 0.10 & 0.15 & 0.20 & \textbf{0.25} \\
 &  & Intolerance & \textbf{0.50} & 0.70 & 0.90 & 0.95 & 0.99 \\ \hline
 & TITE-DC & Selection  $\%$ & \textbf{67.3} & 14.3 & 0.2 & 0 & 0 \\
 &  & No. patients & \textbf{21.2} & 5.0 & 0.5 & 0 & 0 & 13.8 & 20.5 \\
 & TITE-BOIN\textsubscript{DC} & Selection  $\%$ & \textbf{83.1} & 11.9 & 0.1 & 0 & 0 \\
 &  & No. patients & \textbf{23.1} & 6.1 & 0.6 & 0 & 0 & 13.7 & 22.4 \\
 & DC & Selection  $\%$ & \textbf{65.9} & 11.6 & 0 & 0 & 0 \\
 &  & No. patients & \textbf{20.9} & 4.6 & 0.3 & 0 & 0 & 27.7 & 19.1 \\
 & BOIN\textsubscript{DC} & Selection  $\%$ & \textbf{77.1} & 11.3 & 0.1 & 0 & 0 \\
 &  & No. patients & \textbf{22.5} & 5.5 & 0.4 & 0 & 0 & 27.6 & 20.8 \\
 & BOIN & Selection  $\%$ & \textbf{0.6} & 8.6 & 24.6 & 31.3 & 35.0 \\
 &  & No. patients & \textbf{5.0} & 6.8 & 7.4 & 6.0 & 4.9 & 16.0 & 83.4 \\

\end{tabular}%
}
\end{table}

\begin{table}[]
\centering
\label{tb: TITE-DC2}
\resizebox{0.95\textwidth}{!}{%
\begin{tabular}{lllcccccccc} \hline \hline
&  &  & \multicolumn{5}{c}{\textbf{Dose}} & \textbf{Duration} & \textbf{Overdose} \\
  \cmidrule{4-8}
 & \textbf{Design}  &  & $\bm{d_1}$ & $\bm{d_2}$ & $\bm{d_3}$ & $\bm{d_4}$ & $\bm{d_5}$ & \textbf{(Months)} &  \textbf{(\%)} \\ \hline
&  &  & \multicolumn{5}{c}{\textbf{Scenario 5}} \\ 
 &  & DLT & 0.10 & 0.15 & 0.20 & \textbf{0.25} & 0.30 \\
 &  & Intolerance & 0.10 & 0.30 & \textbf{0.50} & 0.70 & 0.90 \\ \hline
 & TITE-DC & Selection  $\%$ & 3.8 & 25.3 & \textbf{56.2} & 13.7 & 0.5 \\
 &  & No. patients & 3.8 & 8.9 & \textbf{8.9} & 3.2 & 0.3 & 16.7 & 11.7 \\
 & TITE-BOIN\textsubscript{DC} & Selection  $\%$ & 5.9 & 40.5 & \textbf{47.9} & 5.5 & 0.1 \\
 &  & No. patients & 8.9 & 12.0 & \textbf{7.3} & 1.6 & 0.2 & 16.1 & 5.9 \\
 & DC & Selection  $\%$ & 4.6 & 27.3 & \textbf{53.7} & 13.0 & 0.8 \\
 &  & No. patients & 8.1 & 9.0 & \textbf{9.6} & 3.0 & 0.3 & 28.6 & 10.9 \\
 & BOIN\textsubscript{DC} & Selection  $\%$ & 6.3 & 38.7 & \textbf{48.2} & 6.6 & 0 \\
 &  & No. patients & 8.5 & 11.8 & \textbf{7.9} & 1.6 & 0.1 & 28.8 & 5.8 \\
 & BOIN & Selection  $\%$ & 5.2 & 23.3 & \textbf{30.7} & 24.0 & 16.6 \\
 &  & No. patients & 7.6 & 8.6 & \textbf{7.0} & 4.3 & 2.5 & 15.9 & 22.6 \\

 &  &  & \multicolumn{5}{c}{\textbf{Scenario 6}} \\ 
 &  & DLT & 0.10 & 0.15 & 0.20 & \textbf{0.25} & 0.30 \\
 &  & Intolerance & 0.30 & \textbf{0.50} & 0.70 & 0.90 & 0.95 \\ \hline
 & TITE-DC & Selection  $\%$ & 17.6 & \textbf{64.8} & 15.1 & 0.5 & 0 \\
 &  & No. patients & 13.3 & \textbf{12.0} & 3.9 & 0.4 & 0 & 15.2 & 14.3 \\
 & TITE-BOIN\textsubscript{DC} & Selection  $\%$ & 23.0 & \textbf{68.8} & 8.1 & 0 & 0 \\
 &  & No. patients & 14.3 & \textbf{12.4} & 3.0 & 0.3 & 0 & 14.9 & 11.1 \\
 & DC & Selection  $\%$ & 16.6 & \textbf{67.6} & 13.4 & 0.6 & 0 \\
 &  & No. patients & 12.6 & \textbf{13.0} & 3.7 & 0.3 & 0 & 28.1 & 13.4 \\
 & BOIN\textsubscript{DC} & Selection  $\%$ & 24.5 & \textbf{65.8} & 9.2 & 0.1 & 0 \\
 &  & No. patients & 13.6 & \textbf{13.2} & 2.9 & 0.2 & 0 & 28.2 & 10.5 \\
 & BOIN & Selection  $\%$ & 5.2 & \textbf{23.4} & 30.5 & 24.0 & 16.7 \\
 &  & No. patients & 7.5 & \textbf{8.6} & 7.0 & 4.3 & 2.5 & 15.9 & 46.0 \\

 &  &  & \multicolumn{5}{c}{\textbf{Scenario 7}} \\ 
 &  & DLT & 0.05 & 0.15 & \textbf{0.25} & 0.35 & 0.45 \\
 &  & Intolerance & 0.30 & \textbf{0.50} & 0.70 & 0.90 & 0.95 \\ \hline
 & TITE-DC & Selection  $\%$ & 14.6 & \textbf{67.5} & 15.0 & 0.7 & 0.1 \\
 &  & No. patients & 12.0 & \textbf{12.9} & 4.3 & 0.4 & 0 & 15.3 & 15.8 \\
 & TITE-BOIN\textsubscript{DC} & Selection  $\%$ & 20.6 & \textbf{71.8} & 7.7 & 0 & 0 \\
 &  & No. patients & 13.1 & \textbf{13.5} & 3.2 & 0.3 & 0 & 14.9 & 11.3 \\
 & DC & Selection  $\%$ & 16.7 & \textbf{68.8} & 12.6 & 0.3 & 0 \\
 &  & No. patients & 11.6 & \textbf{13.9} & 3.8 & 0.3 & 0 & 28.0 & 13.7 \\
 & BOIN\textsubscript{DC} & Selection  $\%$ & 21.6 & \textbf{69.4} & 8.7 & 0 & 0 \\
 &  & No. patients & 12.5 & \textbf{14.1} & 3.1 & 0.2 & 0 & 28.1 & 11.0 \\
 & BOIN & Selection  $\%$ & 2.6 & \textbf{33.0} & 43.3 & 18.2 & 3.0 \\
 &  & No. patients & 6.3 & \textbf{10.5} & 8.4 & 3.7 & 1.1 & 15.9 & 44.0 \\

 &  &  & \multicolumn{5}{c}{\textbf{Scenario 8}} \\ 
 &  & DLT & 0.05 & 0.15 & \textbf{0.25} & 0.35 & 0.45 \\
 &  & Intolerance & 0.10 & 0.20 & 0.30 & 0.40 & \textbf{0.50} \\ \hline
 & TITE-DC & Selection  $\%$ & 1.5 & 26.9 & \textbf{44.2} & 21.1 & 6.1 \\
 &  & No. patients & 6.2 & 9.5 & \textbf{9.0} & 4.1 & 1.2 & 17.7 & 17.6 \\
 & TITE-BOIN\textsubscript{DC} & Selection  $\%$ & 1.8 & 36.0 & \textbf{46.3} & 14.3 & 1.7 \\
 &  & No. patients & 6.8 & 11.4 & \textbf{8.4} & 2.9 & 0.5 & 16.9 & 11.3 \\
 & DC & Selection  $\%$ & 1.8 & 27.2 & \textbf{42.7} & 23.1 & 5.2 \\
 &  & No. patients & 6.1 & 9.1 & \textbf{9.0} & 4.3 & 1.5 & 29.2 & 19.3 \\
 & BOIN\textsubscript{DC} & Selection  $\%$ & 2.2 & 35.1 & \textbf{45.5} & 15.3 & 2.0 \\
 &  & No. patients & 6.7 & 11.2 & \textbf{8.6} & 3.0 & 0.6 & 29.3 & 12.0 \\
 & BOIN & Selection  $\%$ & 2.4 & 33.1 & \textbf{43.1} & 18.4 & 3.1 \\
 &  & No. patients & 6.3 & 10.5 & \textbf{8.4} & 3.8 & 1.1 & 15.9 & 16.2 \\
 
 \end{tabular}%
}
\end{table}

\begin{table}[]
\centering
\label{tb: TITE-DC3}
\resizebox{0.95\textwidth}{!}{%
\begin{tabular}{lllccccccccc} \hline \hline
&  &  & \multicolumn{5}{c}{\textbf{Dose}} & \textbf{Duration} & \textbf{Overdose} \\
  \cmidrule{4-8}
 & \textbf{Design}  &  & $\bm{d_1}$ & $\bm{d_2}$ & $\bm{d_3}$ & $\bm{d_4}$ & $\bm{d_5}$ & \textbf{(Months)} &  \textbf{(\%)} \\ \hline
&  &  & \multicolumn{5}{c}{\textbf{Scenario 9}} \\ 
 &  & DLT & 0.15 & \textbf{0.25} & 0.35 & 0.45 & 0.55 \\
 &  & Intolerance & 0.10 & 0.20 & 0.30 & 0.40 & \textbf{0.50} \\ \hline
 & TITE-DC & Selection  $\%$ & 32.1 & \textbf{42.8} & 19.5 & 3.2 & 0.1 \\
 &  & No. patients & 14.3 & \textbf{9.6} & 4.3 & 1.2 & 0.2 & 16.4 & 19.2 \\
 & TITE-BOIN\textsubscript{DC} & Selection  $\%$ & 27.8 & \textbf{49.0} & 19.5 & 2.2 & 0.4 \\
 &  & No. patients & 13.1 & \textbf{11.1} & 4.5 & 0.9 & 0.1 & 16.1 & 18.4 \\
 & DC & Selection  $\%$ & 29.0 & \textbf{43.9} & 19.0 & 4.3 & 0.5 \\
 &  & No. patients & 13.3 & \textbf{9.8} & 4.6 & 1.3 & 0.3 & 29.5 & 21.2 \\
 & BOIN\textsubscript{DC} & Selection  $\%$ & 29.0 & \textbf{48.3} & 18.1 & 2.9 & 0.3 \\
 &  & No. patients & 13.2 & \textbf{11.0} & 4.5 & 0.9 & 0.1 & 29.5 & 18.7 \\
 & BOIN & Selection  $\%$ & 28.2 & \textbf{48.4} & 18.0 & 3.7 & 0.4 \\
 &  & No. patients & 12.9 & \textbf{11.0} & 4.6 & 1.1 & 0.2 & 15.8 & 19.7 \\

&  &  & \multicolumn{5}{c}{\textbf{Scenario 10}} \\ 
 &  & DLT & 0.01 & 0.05 & 0.12 & \textbf{0.25} & 0.37 \\
 &  & Intolerance & 0.05 & 0.10 & 0.20 & \textbf{0.50} & 0.70 \\ \hline
 & TITE-DC & Selection  $\%$ & 0 & 1.7 & 22.5 & \textbf{60.8} & 15.0 \\
 &  & No. patients & 3.9 & 4.4 & 7.7 & \textbf{9.9} & 4.1 & 18.6 & 13.5 \\
 & TITE-BOIN\textsubscript{DC} & Selection  $\%$ & 0.1 & 1.7 & 37.0 & \textbf{58.2} & 3.1 \\
 &  & No. patients & 3.8 & 5.9 & 10.7 & \textbf{8.2} & 1.4 & 18.9 & 4.6 \\
 & DC & Selection  $\%$ & 0 & 1.2 & 24.0 & \textbf{60.6} & 14.3 \\
 &  & No. patients & 3.7 & 5.8 & 10.5 & \textbf{8.6} & 1.4 & 28.9 & 13.7 \\
 & BOIN\textsubscript{DC} & Selection  $\%$ & 0 & 2.1 & 36.0 & \textbf{57.7} & 4.2 \\
 &  & No. patients & 3.7 & 5.8 & 10.5 & \textbf{8.6} & 1.4 & 29.0 & 4.8 \\
 & BOIN & Selection  $\%$ & 0 & 2.1 & 28.6 & \textbf{51.3} & 19.2 \\
 &  & No. patients & 3.7 & 5.4 & 8.6 & \textbf{8.3} & 4.1 & 16.0 & 13.6 \\

 &  &  & \multicolumn{5}{c}{\textbf{Scenario 11}} \\ 
 &  & DLT & 0.05 & 0.12 & \textbf{0.25} & 0.37 & 0.50 \\
 &  & Intolerance & 0.08 & 0.20 & \textbf{0.50} & 0.67 & 0.90 \\ \hline
 & TITE-DC & Selection  $\%$ & 0.6 & 24.1 & \textbf{60.4} & 14.4 & 0.5 \\
 &  & No. patients & 5.8 & 9.2 & \textbf{11.2} & 3.3 & 0.4 & 17.1 & 12.5 \\
 & TITE-BOIN\textsubscript{DC} & Selection  $\%$ & 1.1 & 34.8 & \textbf{60.2} & 4.0 & 0 \\
 &  & No. patients & 6.1 & 12.2 & \textbf{9.9} & 1.7 & 0.1 & 16.5 & 5.9 \\
 & DC & Selection  $\%$ & 1.1 & 24.4 & \textbf{62.3} & 11.5 & 0.6 \\
 &  & No. patients & 5.6 & 9.0 & \textbf{11.7} & 3.3 & 0.3 & 28.7 & 12.1 \\
 & BOIN\textsubscript{DC} & Selection  $\%$ & 1.0 & 35.1 & \textbf{59.2} & 4.8 & 0 \\
 &  & No. patients & 5.9 & 12.1 & \textbf{10.2} & 1.6 & 0.1 & 28.8 & 5.8 \\
 & BOIN & Selection  $\%$ & 0.9 & 25.6 & \textbf{54.1} & 16.8 & 2.7 \\
 &  & No. patients & 5.5 & 10.1 & \textbf{9.6} & 3.9 & 0.9 & 15.9 & 16.2 \\

 \hline \hline
\end{tabular}%
}
\end{table}

\newpage
\begin{figure}
    \centering
    \includegraphics[width=1\linewidth]{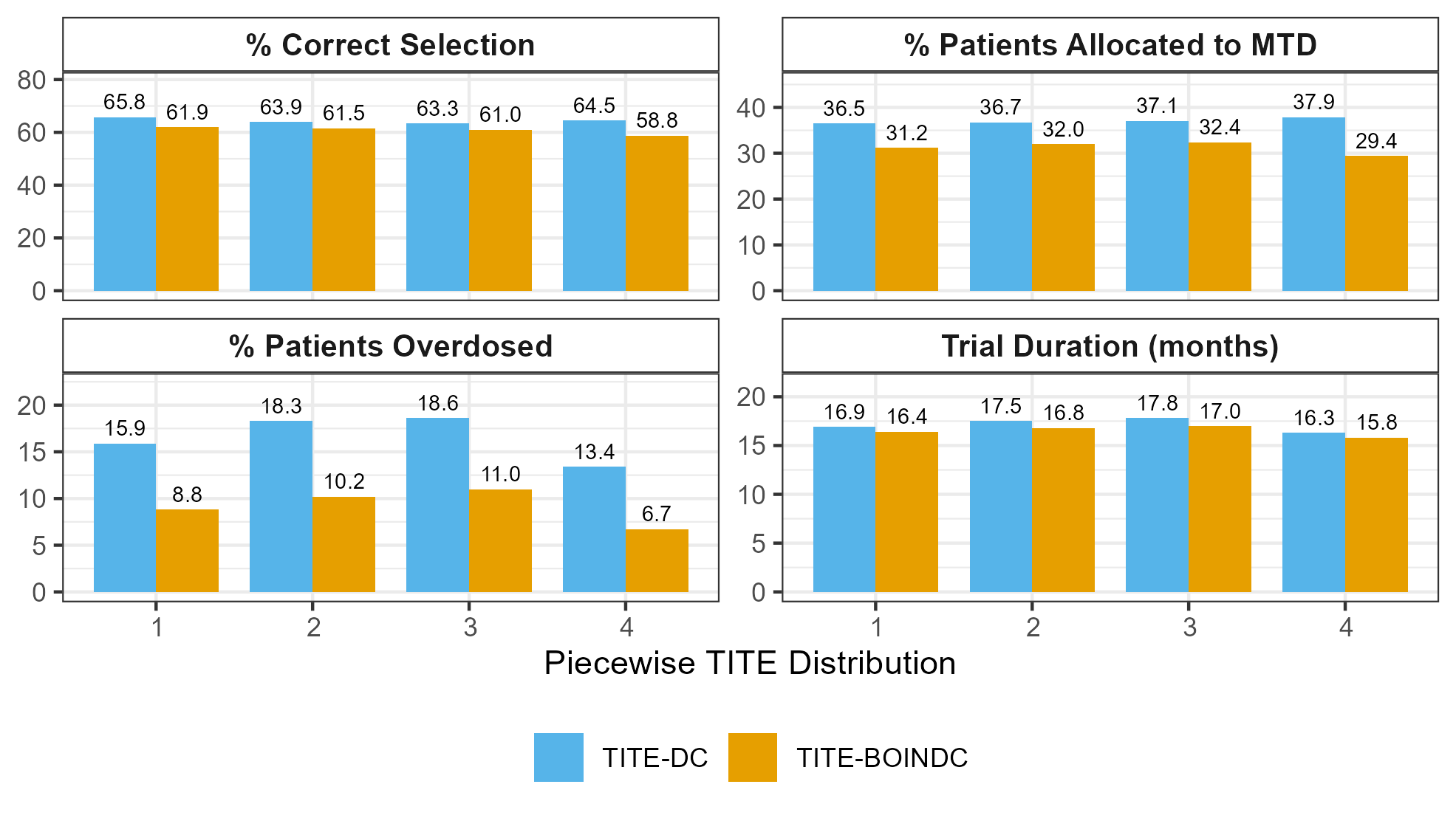}
    \caption{Operating characteristics of TITE-DC and TITE-BOIN\textsubscript{DC} when time to intolerance is weighted 1) uniformly occurring across each treatment cycle, 2) higher towards later cycles, 3) primarily in cycle 3, and 4) primarily in cycle 1.}
    \label{fig:enter-label}
\end{figure}

\begin{figure}
    \centering
    \includegraphics[width=1\linewidth]{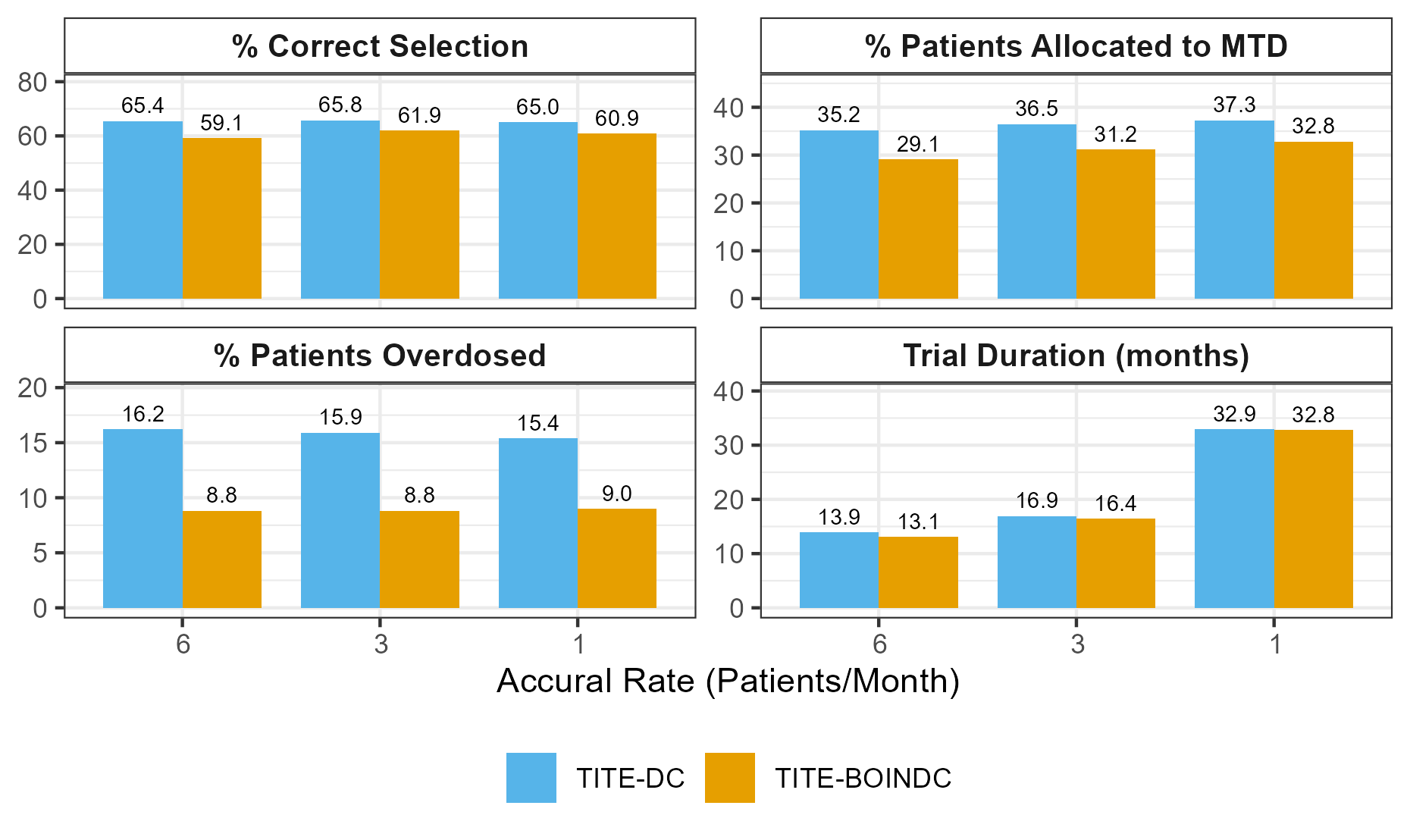}
    \caption{Operating characteristics of TITE-DC and TITE-BOIN\textsubscript{DC} when the accrual rate is varied from 6 patients per month to 1 patient per month.}
    \label{fig:enter-label}
\end{figure}


\newpage
\begin{thebibliography}{99}
\bibitem[Albert and Chib, 1993]{AlbertChib1993}
Albert JH, Chib S. Bayesian Analysis of Binary and Polychotomous Response Data. {\it JASA} 1993;88(422):669-679. doi:10.2307/2290350

\bibitem[Bekele and Thall, 2004]{Thall2004}
Bekele BN, Thall PF. Dose-Finding Based on Multiple Toxicities in a Soft Tissue Sarcoma Trial. {\it JASA} 2004;99(465):26-35. doi:https://doi.org/10.1198/016214504000000043

\bibitem[Chen et al., 2024]{Chen2024}
Chen K, Wang L, Yuan Y. MT-Keyboard: A Bayesian Model-assisted Interval Design to Account for Toxicity Grades and Types for Phase I Trials. {\it SBR} 2024; DOI:10.1080/19466315.2024.2368802. 

\bibitem[Cheung and Chappell, 2000]{Cheung2000}
Cheung YK, Chappell R. Sequential designs for phase I clinical trials with late-onset toxicities. {\it Biometrics} 2000;56(4):1177-1182. doi:10.1111/j.0006-341x.2000.01177.x

\bibitem[CTCAE, 2017]{CTCAE2017}
Common toxicity criteria for adverse events v5.0 (CTCAE). 2017. url:\url{https://ctep.cancer.gov/protocoldevelopment/electronic_applications/ctc.htm#ctc_50}

\bibitem[Ezzalfani et al., 2013]{Ezzalfani2013}
Ezzalfani M, Zohar S, Qin R, Mandrekar SJ, Deley MC. Dose-finding designs using a novel quasi-continuous endpoint for multiple toxicities. {\it Stat Med} 2013;32(16):2728-2746. doi:10.1002/sim.5737

\bibitem[FDA, 2017]{FDA2017}
FDA, Critical Path Institute. Second Annual Workshop On Clinical Outcome Assessment In Cancer Clinical Trials. 2017. url:\url{ https://c-path.org/wp-content/uploads/2017/05/2017_coa_session1consolidatedfinal-.pdf}

\bibitem[Gelman et al., 2008]{Gelman2008}
Gelman A, Jakulin A, Pittau MG, Su YS. A Weakly Informative Default Prior Distribution for Logistic and Other Regression Models. {\it Ann Appl Stat}. 2008;2(4):1360-1383. doi:10.1214/08-AOAS191

\bibitem[Guo et al., 2017]{Guo2017}
Guo B, Yuan Y. Bayesian Phase I/II Biomarker-Based Dose Finding for Precision Medicine with Molecularly Targeted Agents. {\it JASA}. 2017;112(518):508-520. doi: 10.1080/01621459.2016.1228534

\bibitem[Guo and Yuan, 2015]{Guo2015}
Guo B, Yuan Y. A Bayesian dose-finding design for phase I/II clinical trials with non-ignorable dropouts. {\it Stat Med.} 2015;34(10):1721-32.

\bibitem[McKee, 2016]{McKee2016}
Janne PA, Kim G, McKee AE, et al. Dose Finding of Small-Molecule Oncology Drugs: Optimization throughout the Development Life Cycle. {\it Clin Cancer Res}. 2016;22(11):2613-2617. doi:10.1158/1078-0432.CCR-15-2643

\bibitem[Jiang et al., 2024]{Jiang2024}
Jiang L, Yin Z, Yan F, Yuan Y. MC-Keyboard: A Practical Phase I Trial Design for Targeted Therapies and Immunotherapies Integrating Multiple-Grade Toxicities. {\it J Immuno Precis Oncol} 2024. doi:10.36401/JIPO-23-35.

\bibitem[Jin et al., 2014]{Jin2014}
Jin IH, Liu S, Thall PF, Yuan Y. Using Data Augmentation to Facilitate Conduct of Phase I-II Clinical Trials with Delayed Outcomes. {\it J Am Stat Assoc}. 2014;109(506):525-536. doi:10.1080/01621459.2014.881740

\bibitem[Lee et al., 2011]{Lee2011}
Lee SM, Cheng B, Cheung YK. Continual reassessment method with multiple toxicity constraints. {\it Biostat} 2011;12(2):386-398. doi:10.1093/biostatistics/kxq062

\bibitem[Lee et al., 2012]{Lee2012}
Lee SM, Hershman DL, Martin P, Leonard JP, Cheung YK. Toxicity burden score: a novel approach to summarize multiple toxic effects. {\it Ann Oncol} 2012;23(2):537-541. doi: 10.1093/annonc/mdr146

\bibitem[Lin et al., 2018]{Lin2018}
Lin R. Bayesian Optimal Interval Design with Multiple Toxicity Constraints. {\it Biometrics} 2018;74(4):1320-1330. doi:https://doi.org/10.1111/biom.12912

\bibitem[Lin et al., 2020]{BOIN12}
Lin R, Zhou Y, Yan F, Li D, Yuan Y. BOIN12: Bayesian optimal interval phase I/II trial design for utility-based dose finding in immunotherapy and targeted therapies." {\it JCO precision oncology} 2020, 4: 1393-1402.

\bibitem[Liu et al., 2014]{Liu2014}
Liu S, Yin G, Yuan Y. Bayesian Data Augmentation Dose Finding With Continual Reassessment Method And Delayed Toxicity. {\it Ann Appl Stat}. 2014;7(4):1837-2457. doi:10.1214/13-AOAS661

\bibitem[Lin, 2022]{Lin2022}
Liu R, Yuan Y, Lin R, et al. Accuracy and Safety of Novel Designs for Phase I Drug-Combination Oncology Trials. {\it Stat Biopharm Res}. 2022;14(3):270-282. doi:10.1080/19466315.2022.2081602

\bibitem[Mu et al., 2018]{Mu2018}
Mu R, Yuan Y, Xu J, Mandrekar SJ, Yin JY. gBOIN: A Unified Model-Assisted Phase I Trial Design Accounting for Toxicity Grades, and Binary or Continuous End Points. {\it J R Stat Soc Ser C Appl Stat} 2018;68(2):289-308. doi:https://doi.org/10.1111/rssc.12263

\bibitem[Paoletti, 2011]{Paoletti2011}
Postel-Vinay S, Gomez-Roca C, Judson I, et al. Phase I Trials of Molecularly Targeted Agents: Should We Pay More Attention to Late Toxicities? {\it J Clin Oncol}. 2011;29(13):1728-1735. doi:10.1200/JCO.2010.31.9236

\bibitem[Takeda et al., 2023]{Takeda2023}
Takeda K, Yamaguchi Y, Taguri M, Morita S. TITE-gBOIN-ET: Time-to-event generalized Bayesian optimal interval design to accelerate dose-finding accounting for ordinal graded efficacy and toxicity outcomes. {\it Biom J}. 2023;65(7):1-14. doi:10.1002/bimj.202200265

\bibitem[Siu, 2009]{Siu2009}
Tourneau CL, Lee JJ, Siu LL. Dose Escalation Methods in Phase I Cancer Clinical Trials. {\it J Natl Cancer Inst}. 2009;101(10):708-720. doi:10.1093/jnci/djp079

\bibitem[Yuan et al., 2007]{Yuan2007}
Yuan Z, Chappell R, Bailey H. The continual reassessment method for multiple toxicity grades: a Bayesian quasi-likelihood approach. {\it J Biometrics}. 2007;61(1):173-179. doi:10.1111/j.1541-0420.2006.00666.x

\bibitem[Yuan et al., 2016]{Yuan2016}
Yuan Y, Hess KR, Hilsenbeck SG, et al. Bayesian Optimal Interval Design: A Simple and Well-Performing Design for Phase I Oncology Trials. {\it Clin Cancer Res}. 2016;22(17):4291–4301. doi: 10.1158/1078-0432.CCR-16-0592

\bibitem[Yuan et al., 2018]{Yuan2018}
Yuan Y, Lin R, Li D, et al. Time-to-Event Bayesian Optimal Interval Design to Accelerate Phase I Trials. {\it Clin Cancer Res}. 2018;24(20): 4921–4930. doi: 10.1158/1078-0432.CCR-18-0246

\bibitem[Yuan et al., 2024]{Yuan2024}
Yuan Y, Zhou H, Liu S. Statistical and practical considerations in planning and conduct of dose-optimization trials. {\it SAGE}. 2024;21(3):273-286. doi:10.1177/17407745231207085 

\bibitem[Zhou et al., 2022]{Zhou2022}
Zhou Y, Lin R, Yuan Y. TITE-BOIN12: A Bayesian phase I/II trial design to find the optimal biological dose with late-onset toxicity and efficacy. {\it Stat Med}. 2022;41(11):1918-1931. doi:10.1002/sim.9337


\bibitem[Zhou et al., 2018]{Zhou2018}
Zhou H, Yuan Y, Nie L. Accuracy, Safety, and Reliability of Novel Phase I Trial Designs. {\it Clin Cancer Res.} 2018; 24(18):4357-4364. 

\bibitem[Theoret, 2022]{Theoret2022}
Zirkelbach JF, Shah M, Theoret MR, et al. Improving Dose-Optimization Processes Used in Oncology Drug Development to Minimize Toxicity and Maximize Benefit to Patients. {\it J Clin Oncol}. 2022;40(30):3489-3500. doi:10.1200/JCO.22.00371

\end{thebibliography}
\end{document}